\documentclass{article}
\usepackage{amsmath, amssymb, latexsym, epsfig, multicol}
\begin{document}
\newcommand{\di}{\displaystyle}
\newcommand{\reff}[1]{(\ref{eq:#1})}
\newcommand{\labl}[1]{\label{eq:#1}}
\newcommand{\relfactor}[1]{\sqrt{\displaystyle 1 - #1^2/c^2}}
\newcommand{\vc}[1]{{\boldsymbol{#1}}}
\newcommand{\Grad}{\operatorname{grad}}
\newcommand{\Div}{\operatorname{div}}
\newcommand{\Rot}{\operatorname{rot}}

\centerline{\Large\bf The Unipolar Induction}

\vspace{3mm}

\centerline{
P. Hrask\'{o}\footnote{P\'ecsi Tudom\'anyegyetem, 
P\'ecs, Hungary. Email: peter@hrasko.com}}

\vspace{10mm}

\noindent{\large\bf 1 The phenomenon of unipolar induction}

\vspace{5mm}

Consider a permanent magnet in the form of an electrically conducting
disk magnetized along its symmetry axis and rotating around it with
constant angular velocity $\Omega$. If, using sliding contacts, one
connects the axis and the rim of the disk by means of a rigid linear
conductor electric current will go through the latter (unipolar
generator, Fig. 1). An electromotive force arises even when the disk is
made of a dielectric not supporting electric current. The mechanism
which gives rise to this electromotive force is called {\em unipolar
induction}\footnote{See e.g. J. Djuri\'{c} {\em J. Phys. D.} {\bf 9},
2623 (1976) and references to earlier works there.}. 
\vspace{2mm}

\begin{center}
\epsfig{file=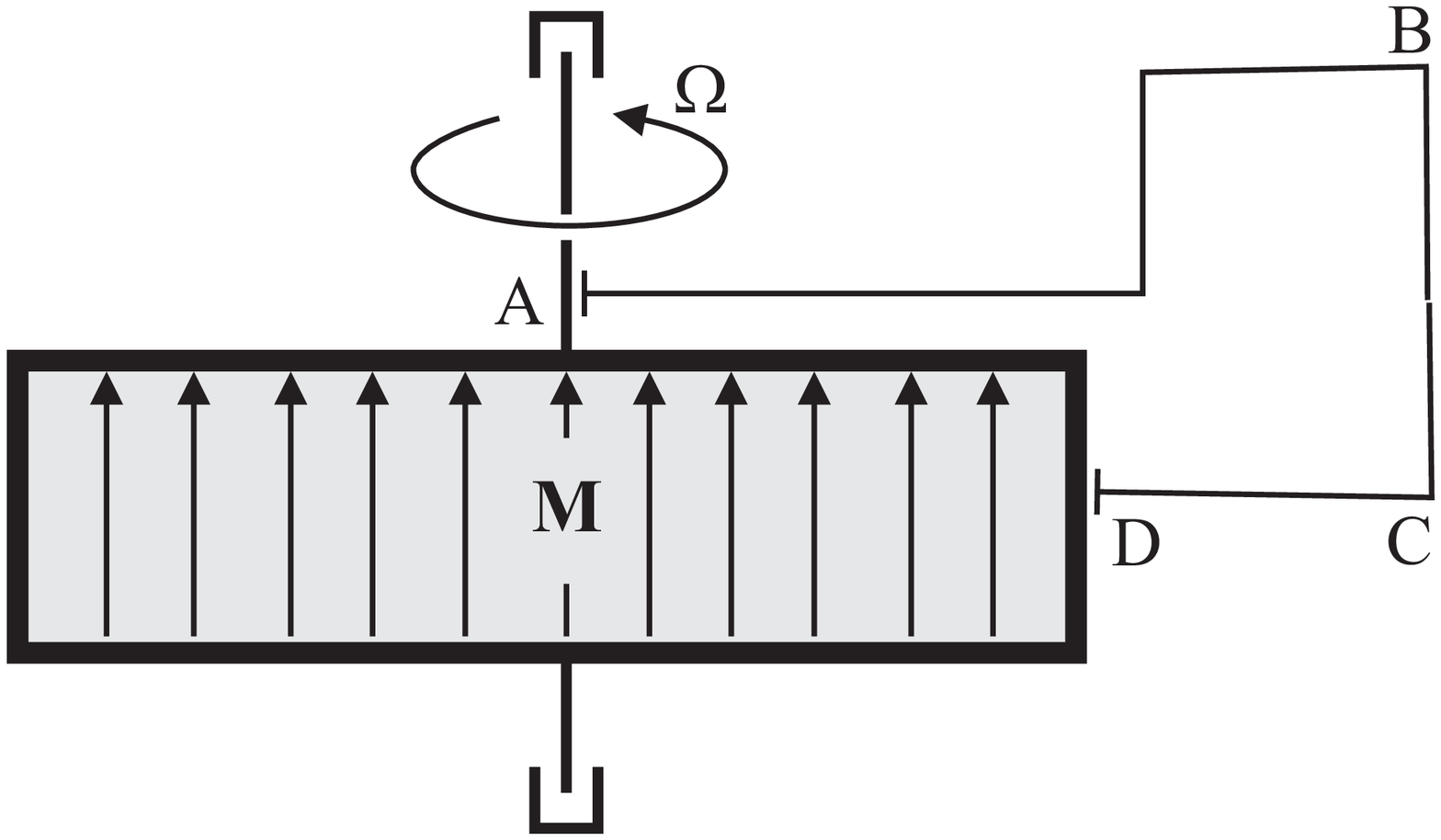,width=60mm}\\
\hspace{-5mm}Figure 1
\end{center}

\vspace{2mm}

How to calculate the magnitude of the effect? When the disk is at rest
and the conductor rotates in the opposite direction with angular
velocity $-\Omega$, it crosses contiuously the lines of force of the
magnet and, therefore, an electromotive force 
\begin{equation} 
{\cal E}_{AD} = \int_{ABCD}\;\bigl [\bigl ((-\vc{\Omega})\times \vc{r}\bigr
)\times\vc{B}\bigr ]\cdot d\vc{r}\labl{erroneous} 
\end{equation} 
arises
in it through motional induction. Since the {\em relative motion} is the
same in both cases one might suppose that this same electromotive force
arises also when the disk rotates together with the lines of force
attached to it and the conductor remains at rest\footnote{L.D. Landau
and E.M. Lifshitz {\em Electrodynamics of Continuous Media} Pergamon
Press 1984, p. 220.}.

This reasoning is, however, of highly dubious value for at least two
reasons. 

First, {\em rotation is
absolute} and it is by no means immaterial whether the disk or the
conductor is rotating. In unipolar
induction the rotation of the disk is supposed to
take place in an inertial frame of reference. When,
on the other hand, the disk is assumed resting,
the point of view is changed and 
the reference frame
becomes rotating.
In mechanics a change of this kind is known to be
accompanied by a corresponding change of the equations of motions,
consisting in the inclusion of centrifugal and Coriolis forces. There is
absolutely no reason to expect that Maxwell-equations remain intact
under the same transformation (See Section 3 below).

Second, the lines of force are of a purely mathematical device, they
motion can be neither observed nor calculated. It is, therefore,
meaningless to differentiate, by reference to the motion of the lines of
force, between the magnetic fields of magnets which rotate with
different angular velocity around their symmetry axis. Hence in the two
cases compared above the magnetic field is actually the same and the
linear conductor which was originally at rest is simply replaced by a
rotating one. These two situations are, however, completely unrelated
since they differ from each other more than in a mere change of the
point of view.

The correct explanation of the phenomenon is based on the observation,
that the elements of the magnet rotate in the field of all the other
elements. The magnet, therefore, rotates in its own magnetic field and
--- due to motional induction --- an
electromotive force
\[(\vc{V}\times\vc{B}) = 
\bigl [(\vc{\Omega}\times
\vc{r})\times\vc{B}\bigr ]\]
will act in the element of it which at the given moment is in the
position $\vc{r}$. Since the magnet is assumed conducting, the free
charges get moved by this force until the field $\vc{E}$ due to the volume
and surface polarization charges compensates the induced electromotive
force within the magnet:
\begin{equation}
\vc{E} = -(\vc{V}\times\vc{B}).\labl{einside}
\end{equation}
It is {\em the electric field of these polarization charges}
outside the magnet
which drives the current in the contour $ABCD$.
Therefore, instead of \reff{erroneous}, the correct electromotive force
is given by the formula
\begin{equation}
{\cal E}_{AB} = \int_{ABCD}\:\vc{E}\cdot d\vc{r}.\labl{correct}
\end{equation}
The electric field, surrounding the magnet, remains obscured in the previous
(erroneous) explanation\footnote{Since the polarization charges rotate
together with the magnet they generate a constant magnetic field which
modifies the original field of the magnet. It is easily seen that this
secondary
field is proportional to $\Omega^2$. In what follows we will confine
ourselves to first order in $\Omega$ in which this modification of the
magnetic field plays no role.}.
But, in spite of their fundamental difference, 
{\em both the correct and the erroneous 
approaches lead to exactly the same electromotive force}.
This follows from the general form\footnote{The proof of this 
formula is found in Appendix 1.} 
\begin{equation}
\dot\Psi = -\oint_{C_t}\;\bigl [\vc{E} + (\vc{V}\times\vc{B})\bigr
]\cdot d\vc{r}\labl{ind}
\end{equation}
of the law of induction which is valid even for moving contours
$C_t$.

Assume that the contour $ABCD$ rotates together with the magnet and
close it by adding a part $DPA$ within the
magnet. Due to axial symmetry the magnetic flux through this closed
contour is constant in time, therefore
\[\oint_{ABCDPA}\;\bigl [\vc{E} + (\vc{V}\times\vc{B})\bigr
]\cdot d\vc{r} = 0.\]
However, as a consequence of \reff{einside}, the integrand on 
the part $DPA$ vanishes 
and we have
\begin{equation}
\begin{gathered}
\int_{ABCD}\; \vc{E}\cdot d\vc{r} = -
\int_{ABCD}\;(\vc{V}\times\vc{B})\cdot d\vc{r} =\\
= - \int_{ABCD}\;\bigl [\bigl (\vc{\Omega}\times
\vc{r}\bigr )\times\vc{B}\bigr ]\cdot d\vc{r}
\end{gathered}\labl{emf}
\end{equation}
\noindent which proves the numerical equality 
of the two forms of ${\cal E}_{AB}$. 

The first method, however, is by no means justified by this coincidence.
When, for example, the magnet is a dielectric the first method leads 
obviously to the same electromotive force as before while the correct
electromotive force given by the second method turns out to depend on
the dielectric constant of the disk (see next section).

Summing up: Unipolar induction consists in the electric polarization of
the permanent magnet, rotating around its symmetry axis. In the special
case of magnets made of conducting material the formula for the
potential drop along contours like $ABCD$ coincides with the formula for
the electromotive force induced in a contour rotating around a magnet
which is at rest in an inertial frame of reference. In a practical
calculation however, instead of \reff{emf},
one can start from the known inside electric field \reff{einside} and
calculate the outside field from the continuity of the tangential
component of $\vc{E}$ across the surface of the magnet and the absence
of the net charge on it. For a
spherical magnet of radius $a$ and constant magnetic dipole density $M$
this procedure leads to the potential\footnote{See paper {\em The rotating
magnet} on my homepage www.hrasko.com/peter .}
\begin{equation}
\Phi = -\frac{2}{9}a^5\mu_0M\Omega\frac{P_2(\cos\vartheta
)}{r^3}\qquad (r\ge a)\labl{potin}
\end{equation}
outside the sphere.

\vspace{5mm}

\noindent{\large\bf 2 The role of the motionally induced 
electric dipole density}

\vspace{5mm}

According to relativity theory
the magnetic dipole density $\vc{M}$, moving with velocity $\vc{V}$,
acquires electric dipole density $\vc{P}_M$. To linear
order in $\Omega$
\[\vc{P}_M = \frac{1}{c^2}(\vc{V}\times\vc{M}).\]
Does this electric polarization
contribute to unipolar induction?

It does, but if the magnet conducts electricity, the consideration of
the previous section remains unaltered. 
In stationary rotation the electric current density $\vc{J}$ 
vanishes in both conductors and dielectrics. According to
Ohm's law we have 
$\vc{J} = \sigma\bigl (\vc{E} + (\vc{V}\times\vc{B})\bigr )$. 
In conductors $\sigma\not= 0$, the condition $\vc{J}=0$ leads to
the electric field \reff{einside} inside the magnet 
and thereby the field outside the magnet becomes unambiguously
determined. It is of no
significance whether this field is generated 
solely by the free charges or
the polarization charge density $ -\Div\vc{P_M}$ also contributes to it.

When, on the other hand, the magnet is made of dielectric, in
the Ohm's law we have $\sigma =0$ and $\vc{J}$ vanishes without requiring
\reff{einside} to be fulfilled. In this case the relevant electrostatic
problem can be solved$^{\footnotesize 6}$, starting from the equation 
$\Div\vc{D} = 0$ where now
\[\vc{D} = \varepsilon\vc{E} - \varepsilon_0\chi (\vc{V}\times\vc{B}) -
\vc{P}_M.\]
The second term on r.h.s. is the polarization
generated by the motionally induced electromotive force ($\chi$ is the
electric susceptibility).

We arrive again at a well defined electrostatic problem which is easily
solved for a spherical magnet. For the potential difference between the
axis of rotation and the "equator" the formula
\[\Delta\Phi^{ins} = \Delta\Phi^{cond}\cdot\frac{2\varepsilon +
\varepsilon_0}{2\varepsilon + 3\varepsilon_0}\]
is obtained ($\Phi^{cond}$ is given in \reff{potin}).
As $\varepsilon\longrightarrow\infty$ we have 
$\Delta\Phi^{ins}\longrightarrow\Delta\Phi^{cond}$ as expected. When 
the magnet is neither conducting nor polarizable we obtain
$\di\Delta\Phi^{ins}_{\varepsilon =\varepsilon_0} =
\frac{3}{5}\Delta\Phi^{cond}$ which is the contribution of $\vc{P}_M$
alone to unipolar induction.

\vspace{5mm}

\noindent{\large\bf 3 Transition to the corotating frame}

\vspace{5mm}

Reference frame must always be chosen so as to simplify the analysis of
the problem under study. Since there are important problems which are
most conveniently studied in the rest frame of the rotating magnet we
turn to the description of the unipolar induction in this latter frame.
Observable properties  such as
whether an electric bulb in the contour $ABCD$ of Fig. 1 
is or is not gleaming 
cannot, of course,
be altered by mere change of the point of view but to verify this we
must first transcript Maxwell-equations into rotating frame.

In order to anticipate the necessary modifications assume that in the
immediate vicinity of the axis of rotation a constant linear charge density of
magnitude $\lambda$ is concentrated. Then, if the equation
$\Div\varepsilon_0\vc{E}=\rho$ remained valid in the rotating frame too, 
the flux of $\vc{E}$ through the multitude of cylindrical surfaces, 
surrounding the
axis of rotation, would remain equal to $\lambda /\varepsilon_0$ per unit
height in both rotating and inertial frames and this requirement would
strongly limit the possible change of the electric field in the
transition from an inertial frame to the corotating one.

But such a change must take place. The Coulomb force $e\vc{E}$ by which the charge
density $\lambda$ acts on a point charge $e$ is certainly different when
the charge is at rest with respect to the one or the other frame. This
is suggested by the transformation law
\[\vc{E}' \approx \vc{E} + (\vc{V}\times\vc{B})\qquad (V \ll c)\]
between two inertial systems. Though inapplicable when transition takes
place
to rotating frame, this equation suggests that the electric field
does change in the latter transition also, leading
an observable difference in the value of the Coulomb
force in the two frames.

Let us assume now that an electric current $I$ flows along the axis of
rotation. Than, if the equation $\Rot\vc{B} = \mu_0\vc{J}$ remained valid
in the rotating frame, the field $\vc{B}$ would presumably also have to
remain essentially the same. But this would almost certainly 
be in conflict with the relativistic
transformation properties of the fields and the expected modification of
the Lorentz-force.

The way out of this dilemma is suggested by the electrostatics of polarized
media: On the level of field equations
the fields $\vc{E}$, $\vc{B}$ which determine the force exerted
on charged particles must be distinguished from the fields $\vc{D}$,
$\vc{H}$ generated by the latters. At the same time, an appropriate pair
of "material equations" must be postulated in order to restore the
correct number of degrees of freedom. The whole process 
must naturally be based 
on the known relativistic transformation properties of the field
in the form of an antisymmetric field tensor $F^{ij}$.

The shortest way to this form of electrodynamics in a rotating
frame goes through general relativity where the problem has been solved in
complete generality for an arbitrary system of coordinates\footnote{See 
L.D. Landau and E.M. Lifshitz {\em The Classical Theory of Fields}
Fourth Revised English Edition, \S\ 90, p. 275, and Appendix 2 at the end
of the present work.}. Specification of the general formulae leads to the
following equations valid in rotating coordinate system:

\begin{equation}
\begin{aligned}
\Div\vc{B} &= 0\\[3mm]
\Div\vc{D} &= \rho
\end{aligned}\qquad
\begin{gathered}
\frac{\partial\vc{B}}{\partial t} + \Rot\vc{E} = 0\\[3mm]
\frac{\partial\vc{D}}{\partial t} = \Rot\vc{H} - \vc{J}.
\end{gathered}
\labl{me}
\end{equation}
The "material equations" are
\begin{gather}
\vc{D} = \frac{\varepsilon_0}{\relfactor{V}}\:\vc{E} +
\frac{1}{c^2}(\vc{H}\times\vc{V})\labl{me1}\\
\vc{B} = \frac{\mu_0}{\relfactor{V}}\:\vc{H} +
\frac{1}{c^2}(\vc{V}\times\vc{E}).\labl{me2}
\end{gather}
$\vc{V}(\vc{r})$ is a vector field present in the rotating frame given
by the formula
$\vc{V} = (\vc{r}\times\vc{\Omega})$. In the rotating frame 
Lorentz-force acquires an additional term:
\begin{equation}
\vc{F}_L = e\vc{E} + e(\vc{v}\times\vc{B})-
\frac{e}{c^2}(\vc{E}\cdot\vc{v})\:\vc{V}.\labl{lf}
\end{equation}

Apply now these equations to unipolar induction.
To this end we have to replace $\vc{H}$ in \reff{me2} by the sum 
$\vc{H} + \vc{M}$ within the magnet.
The jump conditions on the surface of the magnet 
follow from \reff{me} and are the usual
ones: Normal components of $\vc{D}$ and $\vc{B}$ and tangential
components of $\vc{E}$ and $\vc{H}$ are continuous which implies the
continuity of the electric potential.

The same argument which in Section 1 led to the relation \reff{einside} now
gives
$\vc{E} = 0$
inside the magnet. 
Outside we have $\Div\vc{D}=\rho =0$ which leads to the
equation
\begin{equation}
\Div\left (\frac{\varepsilon_0}{\relfactor{V}}\:\vc{E}\right ) = 
-\frac{1}{c^2}\Div
(\vc{H}\times\vc{V}).\labl{diveq}
\end{equation}
This formula is the starting point of the calculation of the electric field in
the corotating frame.

As we have mentioned earlier$^4$ our considerations in the inertial
frame were implicitely restricted to linear order in $\Omega$ since back
reaction of the induced currents on the magnetic field was neglected. In
the rotating frame no such currents arise and nonlinearity in $\Omega$
is made explicite in the equations \reff{me}-\reff{me2} which is a
definite advantage of the rotating frame. 

In order to confine ourselves to linear order again\footnote{This
limit can be directly contrived without making the detour through
general relativity 
(see Landau and Lifshitz {\em Electrodynamics of Contiuous
Media} \S\ 76).} we note that
when $\Omega =0$ the electric field $\vc{E}$ obviously disappears and so
it is proportional to $\Omega$. Since the same is true
for the field $\vc{V}(\vc{r})$ the second term on the r.h.s. of
\reff{me2} is of the order of $\Omega^2$ and must be neglected.
At the same time the square roots in the "material equations" 
and in \reff{diveq} must be replaced by 1.

In this approximation, therefore, the magnetic field in the corotating
system is exactly the same well known dipole field 
as in the laboratory (inertial) frame\footnote{In the literature on the
magnetic hydrodinamics of rotating fluids this fact is often expressed
as e.g. "rigid body rotation... rotates a magnetic field without
distortion" (H. K. Moffat {\em Magnetic Field Generation in Electrically
Conducting Fluids} Cambridge 1978, p. 53)}. It can be substituted into the
r.h.s. of \reff{diveq} which can than be used to calculate $\vc{E}$. 
Since the fields are static we have from \reff{me} $\Rot\vc{E}=0$ and 
an electric potential $\Phi '$ may be introduced 
into \reff{diveq} by the usual relation $\vc{E} = -\Grad\Phi
'$:
\begin{equation}
\triangle\Phi ' = \frac{1}{c^2}\Div (\vc{H}\times\vc{V}).\labl{poteq}
\end{equation}
This equation must be solved under the following conditions: (1) Since
the tangential component of the electric field
on the surface of the magnet is continuous and,
therefore, equal to zero $\Phi '$ must be constant along this surface;
(2) Since the magnet is uncharged the flux of the
field $\vc{D}$ across the closed surfaces, surrounding the magnet, must be
zero. For the spherical magnet the
solution is
\begin{equation}
\Phi '= \Phi - \frac{1}{3}a^3\mu_0M\Omega\:\frac{\sin^2\vartheta}{r}
\qquad (r\ge a)\labl{potrot}
\end{equation}
which is to be compared with \reff{potin}.

Let us calculate finally the electromotive force 
${\cal E}_{AB}'$ in the contour $ABCD$
of Section 1 which rotates with angular velocity $\omega = -\Omega$ in the
corotating system. This force is determined by the Lorentz-force
equation \reff{lf} in which now $\vc{v} = 
(\vc{\omega}\times\vc{r})=
-(\vc{\Omega}\times\vc{r})$.
The third term of the force is of the order of $\Omega^2$ and must be
neglected while the first term's contribution vanishes since the
endpoints of the contour lie on the surface of the magnet which is
equipotential. We have, therefore
\[{\cal E}_{AB}' = 
\int_{ABCD}\;\bigl [\bigl (
(-\vc{\Omega})\times(\vc{r})\bigr )\times\vc{B}\bigr ]\cdot d\vc{r}\]
which is exactly the same expression as \reff{erroneous} and \reff{emf}. When
the magnet is insulating its surface ceases to be equipotential and the
first term of \reff{lf} also contributes to the electromotive force.
This contribution is determined directly by $\vc\Omega$, being independent
of $\vc{v}$.

\vspace{5mm}

\noindent{\large\bf 4 Application of the theory to the Earth}

\vspace{5mm}

Erth's core is a huge rotating magnet and it is tempting to apply the ideas of
the preceding sections to it. To start with let us assume that (1) the
core is a permanent magnet with nonzero conductivity\footnote{The
Glatzmaier-Roberts model of the geomagnetic field 
(G.A.Glatzmaier and
P.H.Roberts {\em Contemporary Physics} {\bf 38}, 269 (1997))
would certainly make
possible a much more realistic treatment.}, (2) the magnetic
field outside the core is a pure dipole field, (3) the geographical and
magnetic poles coincide, (4)
 the mantle is electrically neutral $(\mu = \mu_0, \;\;\varepsilon = 
\varepsilon_0 ,\;\;\sigma = 0)$\footnote{Electric polarizability of the mantle
makes the electric field larger. When the radius of the core is half of
that of the Earth, the correction factor is equal to
$(3301\varepsilon_r - 2321)/(670\varepsilon_r+310)$ 
where $\varepsilon_r$ is the
global relative electric permittivity of the mantle.}, 
and finally (5) the atmosphere is absent.

These are mostly very crude assumptions but they permit us to apply the
formulas of the previous sections without further refinements. The electric
field at a given point on the surface of the Earth can be calculated
from \reff{potrot}. Assumption (2) allows to replace the unknown
polarization density $M$ with the empirical value $B$ of the magnetic
field on the Equator, equal to $B=33\:\mu T$. On the Equator
the electric field given
by \reff{potrot} is a purely radial one and is given by the
formula
\[E' = \Omega R_c\left (\frac{R}{R_c} - \frac{R_c}{R}\right )\cdot B\]
where $R$ and $R_c$ are the Earth's and the core's radiuses
respectively. Substituting $R_c\approx R/2\approx 3.10^6\:m$, $\Omega =
2\pi /24\: h^{-1} = 7.3\cdot 10^{-5}\:s^{-1}$ and $B=33.10^{-6}\:T$ we obtain
$E\approx 10\;mV/m$ which is in itself of a measurable  magnitude but 
stronger fields of atmospheric origin make its observation rather
problematic.

In march of 1996 a spherical 1.6-meter diameter satellite was
released out from the payload bay of Space Shuttle Columbia
during its orbiting at a height about $h$=90 km above the Earth. Its tether
a long conducting cable (up to the length of 19.1 km)
of resistance $R$, served (among others) to 
measure the radial electric field (called {\em ambient field}) in the
vicinity of the orbit. The system is conveniently called Tethered
Satellite System (TSS).

Free electrons in the ionosphere where TSS
operated were attracted to the satellite. The electrons
travelled along the tether to the orbiter, producing the current $I$.
The electric circuit was
closed by means of an electron generator on the orbiter which returned
charged particles back into the ionosphere. The product $IR$ 
gave the electromotive force along the tether the dominant part of which
was due to the motional induction in the magnetic field of the 
Earth\footnote{Since this electromotive force is of outward direction (opposite
to the Earth) the orbiter had to revolve nearer to the Earth.} and
could be calculated with sufficiently high accuracy. From the rest of the
electromotive force the ambient field could be estimated.

\vspace{2mm}

\begin{center}
\epsfig{file=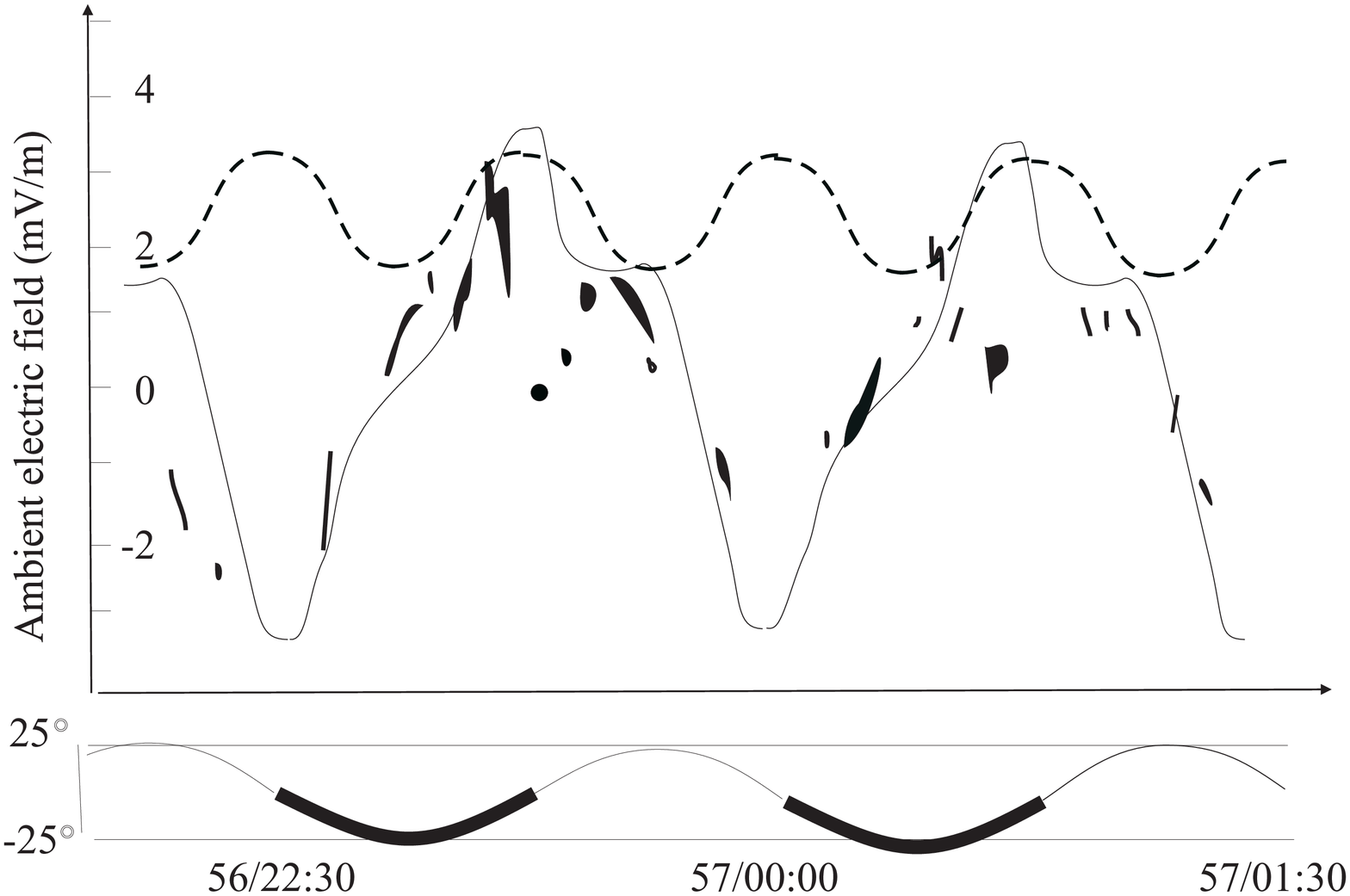,width=60mm}\\
Figure 2
\end{center}

\vspace{2mm}

Results of the measurement\footnote{S. D.
Williams et al. {\em Geophys. Res. Lett.}, {\bf 25} 445 (1998). 
The lower curve shows 
TSS latitude with thick curve indicating local night.} are 
summarized on Fig. 2. The solid line is a
theoretical curve calculated from the theory of "tidally driven neutral
winds" which is outside my competence. The agreement with
observations is considered satisfactory.

The dashed curve added by myself has been calculated according to the formula
\[E_r(t) = -2(\Omega R_c)\:\frac{R_c}{R}\:P_2(-\sin\alpha\cdot\sin\omega
t)\cdot B\]
derived from \reff{potin}; $\alpha$ denotes the angle between 
the orbital and equatorial planes
and $\omega$ is the angular velocity of the satellite system (the radius $R+h$
of the orbit was replaced by $R$). In the actual experiment their values
were $\alpha = 25^\circ$ and $\omega = 2\pi /90\: min^{-1}$. The azimuthal
angle $\omega t$ in the orbital plane is measured from the intersection
with the equatorial plane.

Though the order of magnitude agreement is quite impressive it must not be
taken at face value. The curve was calculated under the assumption
(5) of no atmosphere while the measurement itself was made possible
exclusively thanks to the existence of the ionosphere. What the
order of magnitude of the electric field does probably indicate is that
this field should be taken into consideration among the {\em input data}
when complicated processes in the ionosphere are studied.

One may wonder whether unipolar
induction in the Earth's core might not have some unexpected 
influence on satellites not  
specifically designed to observe the ambient electric field. Flyby
anomaly\footnote{J.D.Anderson et al, {\em Phys. Rev. Lett.} {\bf 100},
091102 (2008).} is a phenomenon which, in spite of the efforts of its
discoverers, remained so far unexplained. In this situation even the
most improbable explanations deserve attention. A suggestion of this kind 
might be an effect due to the electric field of the Earth. 
If satellites can somehow (perhaps by
photoeffect) get electrically charged then the quadrupole character of
this field might account for the latitude dependence of the observed
anomaly.

\vspace{5mm}

\noindent{\large\bf Appendix 1: Derivation of \reff{ind}\footnote{H. K.
Moffat {\it op. cit.} p. 32}}

\vspace{5mm}

Let us start with formula
\begin{equation}
\Psi (t) = \int_\Sigma \;{\bf B}\cdot d\vc{\Sigma} = \oint_{C_t}\;\vc{A}
\cdot d\vc{r}\labl{19b}
\end{equation}
for the magnetic flux $\vc{\Psi}$ through the closed 
material contour $C_t$,
following from the definition $\vc{B} = \Rot\vc{A}$ of the vector
potential. The contour $C_t$ may vary its shape and length in time and
is given in parametric form as 
\begin{gather*}
\vc{r} =
\vc{r}(\lambda ,t)\qquad (0\le \lambda \le 1),\\
\vc{r}(1,t) = \vc{r}(0,t),
\end{gather*}
where it is understood that, during the deformations, the
same value of the parameter $\lambda$ remains attached to the same
material point\footnote{Since the integral on the r.h.s. 
of \reff{ind} is not invariant
under time dependent reparametrization of the contour it is
defined only if the parametrization is given in physical terms (up to an 
arbitrary choice at some given
moment of time).}.
 
Since the contour is closed the vector function
$\vc{r}(\lambda ,t)$ is a periodic function of $\lambda$ of period
$1$: $\vc{r}(\lambda +1,t) = \vc{r}(\lambda ,t)$. 
The velocity of the point at $\lambda$ is equal to $\di\vc{v}
\equiv\dot{\vc{r}} = \frac{\partial\vc{r}(\lambda ,t)}{\partial t}$.

Using this representation of the contour, the integral in \reff{19b} can
be written in the form
\[\Psi (t) = \int_0^1\:A_i\:\frac{\partial
x_i(\lambda ,t)}{\partial\lambda}\:d\lambda,\]
where the notation $x_1,\;x_2,\;x_3 \equiv x,\;y,\;z$ is introduced
and summation over repeated indices is understood.
In the integrand the components 
$A_i\equiv A_i\bigl (\vc{r}(\lambda ,t),\;t\bigr )$ are functions of both
$t$ and $\lambda$.

For the time derivative of the flux we obtain
\begin{equation}
\begin{gathered}
\frac{d\Psi}{dt} = \int_0^1\:\frac{d}{dt}\left (A_i\frac{\partial
x_i}{\partial\lambda}\right )d\lambda =\\
=\int_0^1\left [\frac{dA_i}{dt}\cdot\frac{\partial x_i}{\partial\lambda}
+ A_i\frac{\partial}{\partial t}\left (\frac{\partial
x_i}{\partial\lambda}\right )\right ]\:d\lambda .
\end{gathered}\labl{20b}
\end{equation}
Here
\begin{equation}
\frac{dA_i}{dt} = \frac{\partial A_i}{\partial t} + \frac{\partial
A_i}{\partial x_j}\frac{\partial x_j}{\partial t} =
\frac{\partial A_i}{\partial t} + v_j\frac{\partial
A_i}{\partial x_j}\labl{20a}
\end{equation}
is the "material derivative" of the vector potential on the contour.

In the second term of the integrand
\[\frac{\partial}{\partial t}\left (\frac{\partial
x_i}{\partial\lambda}\right ) =
\frac{\partial}{\partial\lambda}\left (\frac{\partial
x_i}{\partial t}\right ) = \frac{\partial v_i}{\partial\lambda},\]
hence
\begin{equation}
\frac{d\Psi}{dt} = \int_0^1\left (\frac{\partial A_i}{\partial
t}\frac{\partial x_i}{\partial\lambda} + v_j\frac{\partial A_i}{\partial
x_j}\frac{\partial x_i}{\partial\lambda} + A_i\frac{\partial
v_i}{\partial\lambda}\right )\:d\lambda .
\labl{21a}
\end{equation}
The third term can be written as 
\[\di\frac{\partial}{\partial\lambda}(A_iv_i) - \frac{\partial
A_i}{\partial\lambda}v_i,\] 
the first of which does not contribute:
\[\int_0^1\:\frac{\partial}{\partial\lambda}\:(A_iv_i)\:d\lambda =
(A_iv_i)_{\lambda =1} - (A_iv_i)_{\lambda =0} = 0,\]
and so in the integrand of \reff{21a} we have
\[A_i\frac{\partial v_i}{\partial\lambda} = -v_j\frac{\partial
A_j}{\partial\lambda} = -v_j\frac{\partial A_j}{\partial
x_i}\frac{\partial x_i}{\partial\lambda}.\]
The formula \reff{21a} can then be transformd into
\begin{equation}
\frac{d\Psi}{dt} = \oint_{C_t}\:\left [\frac{\partial A_i}{\partial t} +
v_j\left (\frac{\partial A_i}{\partial x_j} - \frac{\partial
A_j}{\partial x^i}\right )\right ]\:dx_i.\labl{21b}
\end{equation}
In the integrand
\[v_j\left (\frac{\partial A_i}{\partial x_j}-
\frac{\partial A_j}{\partial x_i}\right ) =
-(\vc{v}\times\Rot\vc{A})_i ,\]
therefore
\begin{equation}
\frac{d\Psi}{dt} = \oint_{C_t}\:\left [\frac{\partial\vc{A}}{\partial t}
- (\vc{v}\times\Rot\vc{A})\right ]\cdot d\vc{r}.\labl{22a}
\end{equation}
Here
\[\Rot\vc{A} = \vc{B}\quad \text{and}\quad
\frac{\partial\vc{A}}{\partial t} = - \vc{E} - \Grad\phi ,\]
where $\phi$ is the scalar potential. Substituting these into \reff{22a}
the scalar potential gives no contribution and we arrive at the formula
\reff{ind}.

This formula, therefore, is an exact consequence of the
Maxwell-equations alone but it becomes especially useful when, taking into
consideration the Lorentz-force too, we interprete the integral in it as
the total electromotive force, acting in the contour. This
interpretation, however, is valid only in quasistationary approximation
i. e. when the perturbances may be assumed to cross the domain of the
contour instantaneously.

\vspace{5mm}

\noindent{\large\bf Appendix 2: Maxwell-equations in general coordinates}

\vspace{5mm}

Here we summarize the formulae of Landau and Lifshitz$^7$, using SI.

The space-time coordinate system is characterized by the metric tensor
$g_{ij}$ (latin indices take on values $0,\;1,\;2,\;3$). The three
dimensional metric is denoted by $\gamma_{\alpha\beta}$ (greek indices
are of value $1,\;2,\;3$) which in terms of $g_{ij}$ is given by the
formula
\begin{equation}
\gamma_{\alpha\beta} = -g_{\alpha\beta} + hg_\alpha
g_\beta ,\labl{gamma}
\end{equation}
where 
\[h=g_{00},\qquad g_\alpha = -\frac{g_{0\alpha}}{g_{00}}.\]
Further quantities which will be needed are
\begin{gather*}
\gamma = \det\gamma_{\alpha\beta}\qquad g = \det g_{ij} = -h\gamma\\
\gamma^{\alpha\beta} = -g^{\alpha\beta}\qquad
g^\alpha = \gamma^{\alpha\beta}g_\beta = -g^{0\alpha}.
\end{gather*}
We have further
\[g_{ik}g^{kj} = \delta_i^j,\qquad \gamma_{\alpha \sigma}\gamma^{\sigma
\beta} = \delta_\alpha^\beta .\]
The Maxwell-equations are 
\begin{alignat}{2}
\Div\vc{B} &= 0 & \qquad \Rot\vc{E} &=
-\frac{1}{\sqrt{\gamma}}\frac{\partial}{\partial
t}(\sqrt{\gamma}\vc{B})\labl{1.pair}\\
\Div\vc{D} &= 0 & \qquad \Rot\vc{H} &=
\frac{1}{\sqrt{\gamma}}\frac{\partial}{\partial t}(\sqrt{\gamma}\vc{D})
+ \vc{J},\labl{2.pair}
\end{alignat}
in which $\Rot$ and $\Div$ are understood in form valid in general
coordinates:
\begin{equation}
\begin{gathered}
(\Rot\vc{a})^\alpha =
\frac{1}{2\sqrt{\gamma}}\epsilon^{\alpha\beta\gamma}\left (
\frac{\partial a_\gamma}{\partial x^\beta} -
\frac{\partial a_\beta}{\partial x^\gamma}\right ),\\
\Div\vc{a} = \frac{1}{\sqrt{\gamma}}\frac{\partial}{\partial
x^\alpha}(\sqrt{\gamma}\:a^\alpha )
\end{gathered}\labl{rotdiv}
\end{equation}
The "material equations" are
\begin{equation}
\begin{aligned}
\vc{D} &= \frac{\varepsilon_0}{\sqrt{h}}\vc{E} +
\frac{1}{c}(\vc{H}\times\vc{g})\\
\vc{B} &= \frac{\mu_0}{\sqrt{h}}\vc{H} +
\frac{1}{c}(\vc{g}\times\vc{E}).
\end{aligned}\labl{maxw}
\end{equation}
The continuity equation:
\begin{equation}
\frac{1}{\sqrt{\gamma}}\frac{\partial}{\partial t}(\sqrt{\gamma}\rho ) +
\Div\vc{J} = 0,\labl{ce}
\end{equation}
and, finally, the Lorentz-force:
\[\vc{F}_L = e\vc{E} + e(\vc{v}\times\vc{B}) -
\frac{e}{c}(\vc{E}\cdot\vc{v})\:\vc{g}.\]

The connection between 3D and 4D quantities:
\begin{alignat*}{2}
E_\alpha &= cF_{0\alpha} & \qquad B^\alpha &=
-\frac{1}{2\sqrt{\gamma}}\epsilon^{\alpha\beta\gamma}F_{\beta\gamma}\\
D^\alpha &= -\varepsilon_0c\sqrt{h}F^{0\alpha} & \qquad
\mu_0H_\alpha &=
-\frac{\sqrt{-g}}{2}\epsilon_{\alpha\beta\gamma}F^{\alpha\beta}.
\end{alignat*}
From these the components $E^\alpha$, $B_\alpha$, $D_\alpha$ and $H^\alpha$ 
are obtained with the aid of the 3-dimensional metric.

Specialization to rotating coordinates: In cylindrical coordinates
the tranformation formulae from the (primed) inertial system to the (unprimed)
rotating system are
\[t'=t,\quad r'=r,\quad \vartheta '=\vartheta ,\quad \varphi ' = \varphi
+ \Omega t\]
which leads to the line element
\begin{gather*}
ds^2 = \left [1 - \left (\frac{\Omega r\sin\vartheta}{c}\right
)^2\right ]c^2dt^2 - 2\frac{\Omega r^2\sin^2\vartheta}{c}d\varphi\: d(ct)
-\\
- dr^2 - r^2(d\vartheta^2 + \sin^2\vartheta\: d\varphi^2).
\end{gather*}
From this line element the metric tensor $g_{ij}$ (and $h$) can be read off
and we have
\begin{gather*}
g_r = g_\vartheta = 0\qquad g_\varphi = \frac{\Omega
r^2\sin^2\vartheta}{c.h}\\
\gamma_{rr} = 1,\qquad \gamma_{\vartheta\vartheta} = r^2,\qquad
\gamma_{\varphi\varphi} = \frac{r^2\sin^2\vartheta}{h}.
\end{gather*}
Since the gamma-tensor is diagonal we have $\gamma^{\alpha\alpha} =
1/\gamma_{\alpha\alpha}$ and, therefore
\[g^r=g^\vartheta =0,\qquad g^\varphi = \frac{\Omega}{c}.\]

These formulae lead immediately to the equations \reff{me} -\reff{lf} 
in which the space coordinates can be chosen arbitrarily.

\end{document}